\documentclass[prl,aps,preprint,showpacs,showkeys]{revtex4}
\usepackage{graphicx}
\usepackage{dcolumn}
\usepackage{bm}
\usepackage{hyperref}  
\usepackage{latexsym}
\usepackage{color} 
\begin{document}
\date{\today}
\title{A Fractal Space-filling Complex Network}

\author{D. J. B. Soares$^{1,4}$, J. Ribeiro Filho$^{1,2}$,A. A. Moreira$^1$,
 D. A. Moreira$^3$, and G. Corso$^{3,4}$ }

\affiliation{ $^1$
Departamento de F\'{\i}sica, Universidade Federal do Cear\'a, 
60451-970 Fortaleza, CE, Brazil} 
\affiliation{ $^2$
Curso de Matem\'atica, Universidade Estadual Vale do Acara\'u 62040-370, 
Sobral, CE, Brazil}
\affiliation{ $^3$ Departamento de F{\'\i}sica Te\'orica e Experimental,
Universidade Federal do Rio Grande do Norte, Campus Universit\'ario,
59078 970 Natal, RN, Brazil}
\affiliation{ $^4$ Departamento de Biof{\'\i}sica e Farmacologia,
Centro de Bioci\^encias,
Universidade Federal do Rio Grande do Norte, Campus Universit\'ario
59072 970, Natal, RN, Brazil}
 
\hspace{1cm}

\begin{abstract}

We study in this work the properties of the $Q_{mf}$ network which is
 constructed from an anisotropic partition of the square, the multifractal tiling. 
This tiling is build using a single parameter $\rho$, in the limit of 
$\rho \rightarrow 1$ the tiling degenerates into the square lattice that is 
associated with a regular network. 
 The $Q_{mf}$ network is a space-filling network with the  
following characteristics: it shows a power-law distribution of 
connectivity for $k>7$ and it has an high clustering coefficient when compared 
with a random network associated. In addition the $Q_{mf}$ network 
satisfy the relation $N \propto \ell^{d_f}$ where $\ell$ is a typical length 
of the network (the average minimal  distance) and 
$N$ the network size. We call $d_f$ the fractal 
dimension of the network. In the limit case of 
$\rho \rightarrow 1$ we have $d_f \rightarrow 2$. 

\end{abstract}

\pacs{89.75.Da,  89.75.Hc, 61.43.Hv}

\keywords{networks, fractals, space-filling, multifractal tiling, small-world}

\maketitle

\section{I - Introduction}

The last years have seen an increasing interest in network studies 
in  physics \cite{barabasi,strogatz}. Despite graph 
theory have been a research topic of mathematics and science 
of computation, the physicists have driven their attention 
to networks that show distribution of connectivity, $P(k)$, following a 
power law and that present small world effect: $\ell \propto ln N$, for 
$\ell$ a typical network distance and $N$ the size of the network. 
A recent article \cite{nature} points the difficulty to put together these 
two aspects in a common broad scale-free framework, it means, a fractal 
paradigm. In fact, a real fractal should have $N \propto \ell^{d_f}$ for 
$d_f$ the fractal dimension, and not a logarithm dependence between 
$N$ and $\ell$. In this work we present a network that have the following 
characteristics: it  is fractal, $N \propto \ell^{d_f}$, 
it is scale-free $P(k) \propto k^{-\gamma}$ for a large range of $k$, and 
 it is complex in the sense of having large clustering coefficient.  
The scaling relation $N \propto \ell^d$ is trivially fulfilled for 
regular lattices where $d$ is an integer, the topologic dimension of the space. 
The complex fractal network we explore in this paper has a non integer $d_f$.

Recently  networks  embedded in metric spaces 
have been investigated in the literature \cite{Soares} because of 
the large applications of networks that effectively occupy a volume in 
3-dimensions. In addition, motived by microprocessors design 
 space-filling networks \cite{andrade} have been studied. 
We cite these new trends in networks because the distinguished 
network we analyze in this paper 
 is embedded in a metric space and furthermore it is space filling.
In fact our network has a geometrical  inspiration, it comes
 from a partition of the square.  Indeed 
 this network is originated from a singular tiling 
that  has the additional property of being a multifractal partition of the square 
 \cite{corso1}. 

We follow the previous literature 
\cite{corso1,lim,randi1,randi2,multphy} and call this object the 
multifractal tiling, $Q_{mf}$.  
The $Q_{mf}$ tiling was developed in the context of modeling transport 
and percolation in heterogeneous porous media. 
In fact, a broad set of irregular and
heterogeneous systems are model in the literature using a multifractal
approach. We cite systems in geology \cite{Riedi,Hermann},
atmospheric science \cite{Muller1} and chemistry \cite{stan}. 
Oil reservoirs are complex anisotropic structures whose 
treatment have been challenged science and technique because of its 
non trivial geometry. 
Inspired in the description of oil reservoirs it was developed
\cite{corso1} the  $Q_{mf}$ tiling. 
In reference \cite{lim} an exhaustive study of the
percolation threshold of $Q_{mf}$ was performed, in  
\cite{randi1} a random version of  $Q_{mf}$ was created, in
\cite{randi2} some of its percolation  critical exponents were
found, and in \cite{multphy} a numerical study of its coordination 
number is done. 
 
In this work we explore the network properties of 
the $Q_{mf}$, it means its topology, the study 
of the connections (neighborhood)  among the cells of the tiling. 
  The paper is organized as follows. 
In section $2$ we describe in some detail 
the process of construction of the $Q_{mf}$ tiling
 and show its more important properties.
In section $3$ we show the main 
results concerning the properties of the network: 
the distribution of connectivity, the scaling of the  
 the average minimal distance between two sites and  the 
clustering coefficient. 
Finally in section $4$ we present our conclusions,  
discuss the main implications of our results and compare 
the properties of  the $Q_{mf}$ network with other networks in 
the literature. 

\section{II - The multifractal object}

The multifractal tiling is a peculiar partition of the square. 
It is interesting to think about the $Q_{mf}$ tiling in contrast to  
the square lattice.  
The square lattice can be constructed using the following algorithm: 
take a square and cut it symmetrically with horizontal and  vertical 
lines. This procedure produces four square cells. 
Repeat this procedure $n$ times inside each new block  and you have finally
a square lattice with $2^n$ cells. The $Q_{mf}$ object is generated 
in a similar way as the square lattice above described, but instead of using a 
symmetric partition we perform horizontal and vertical sections following  
a given ratio.

\begin{figure} \begin{center}
\includegraphics[width=3cm]{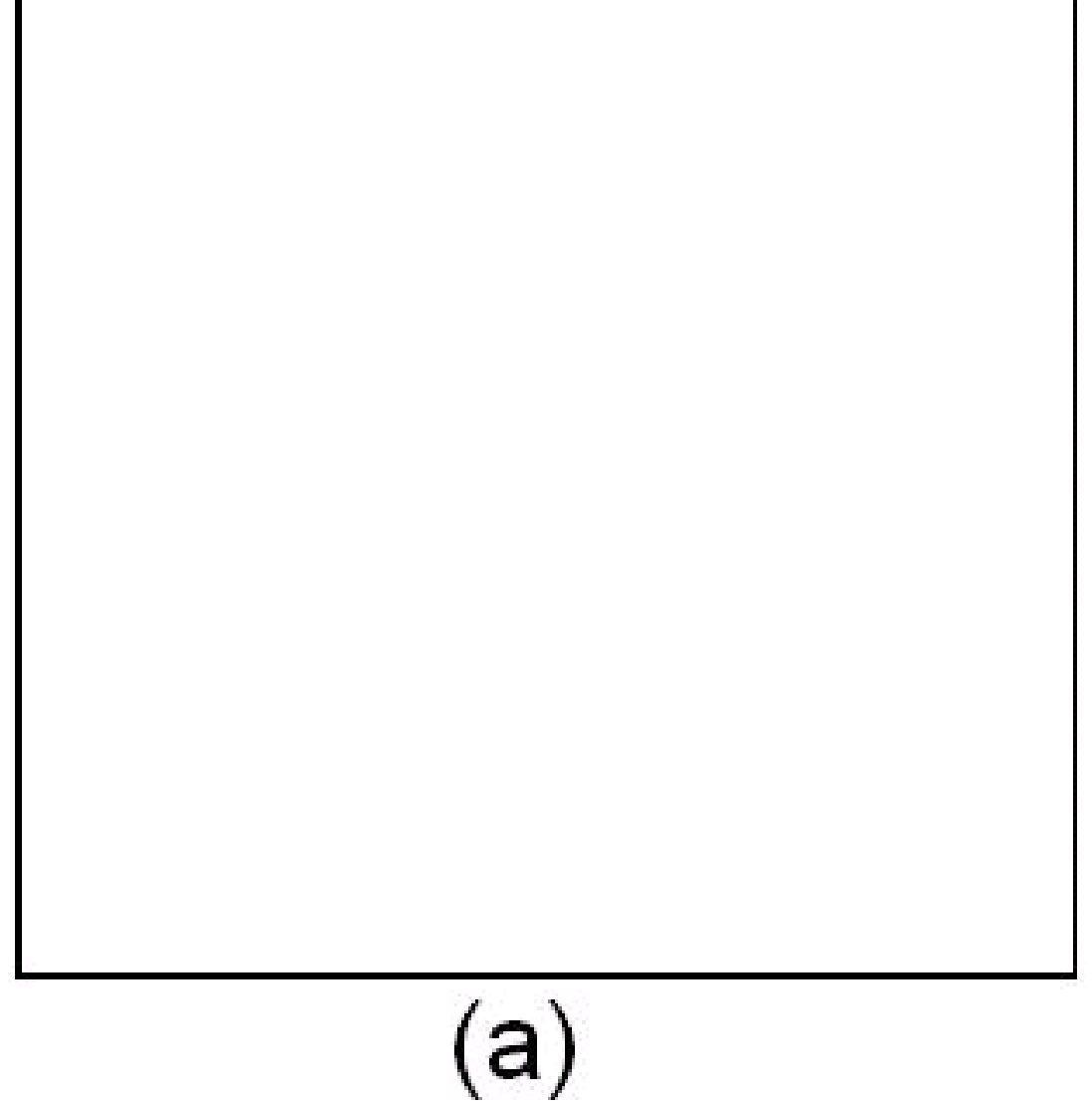}
\includegraphics[width=3cm]{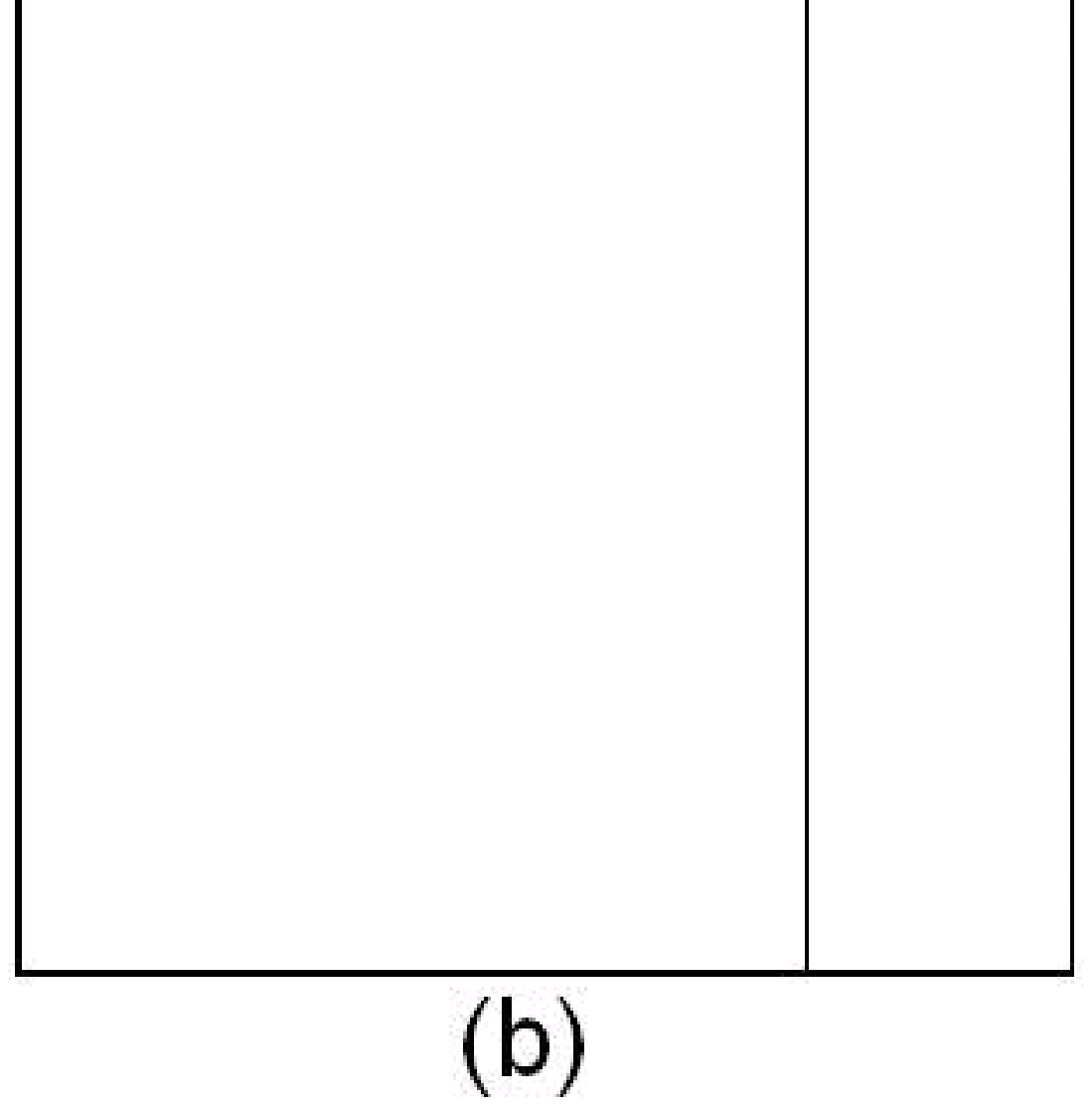}
\includegraphics[width=3cm]{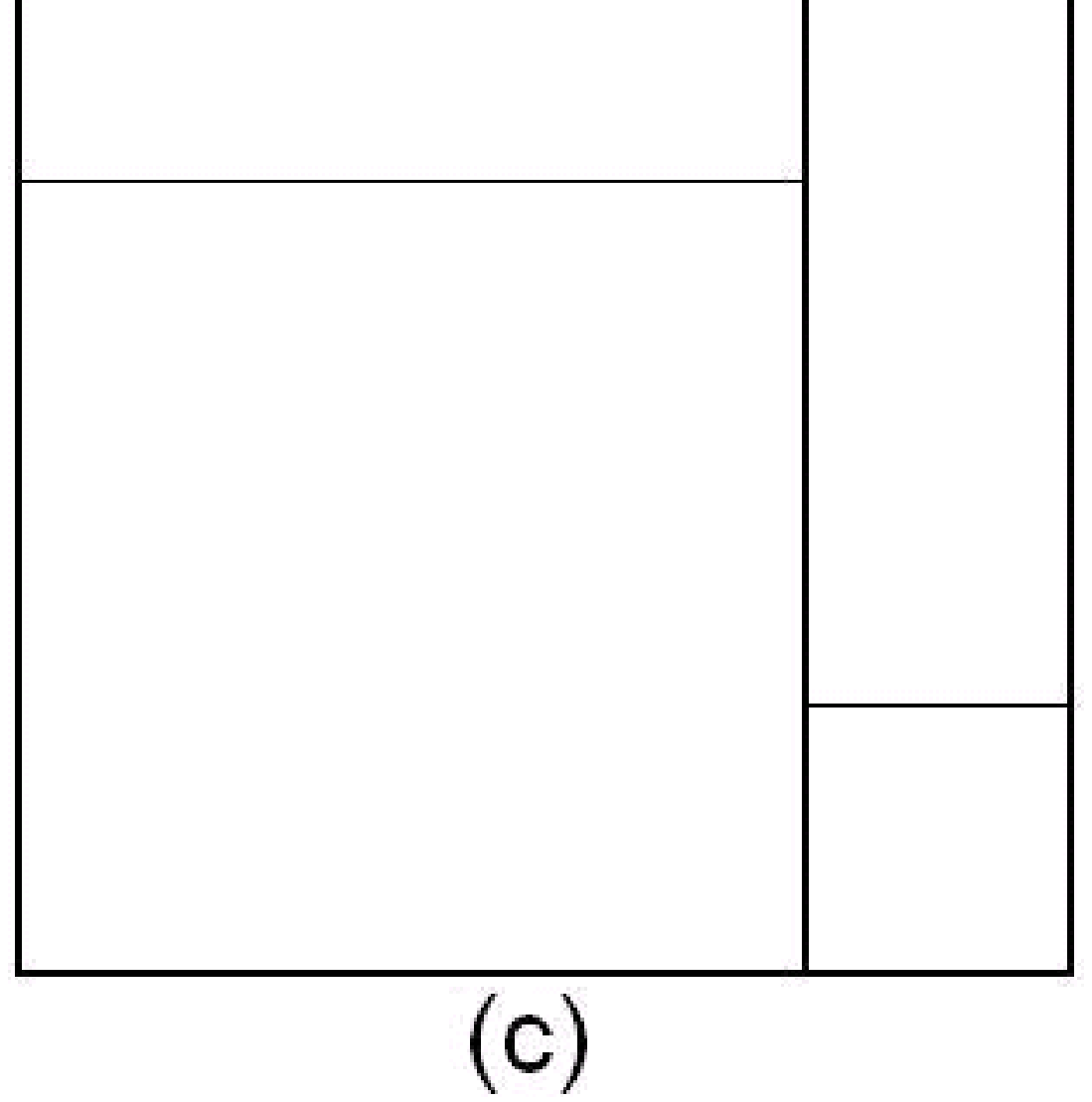}
\includegraphics[width=3cm]{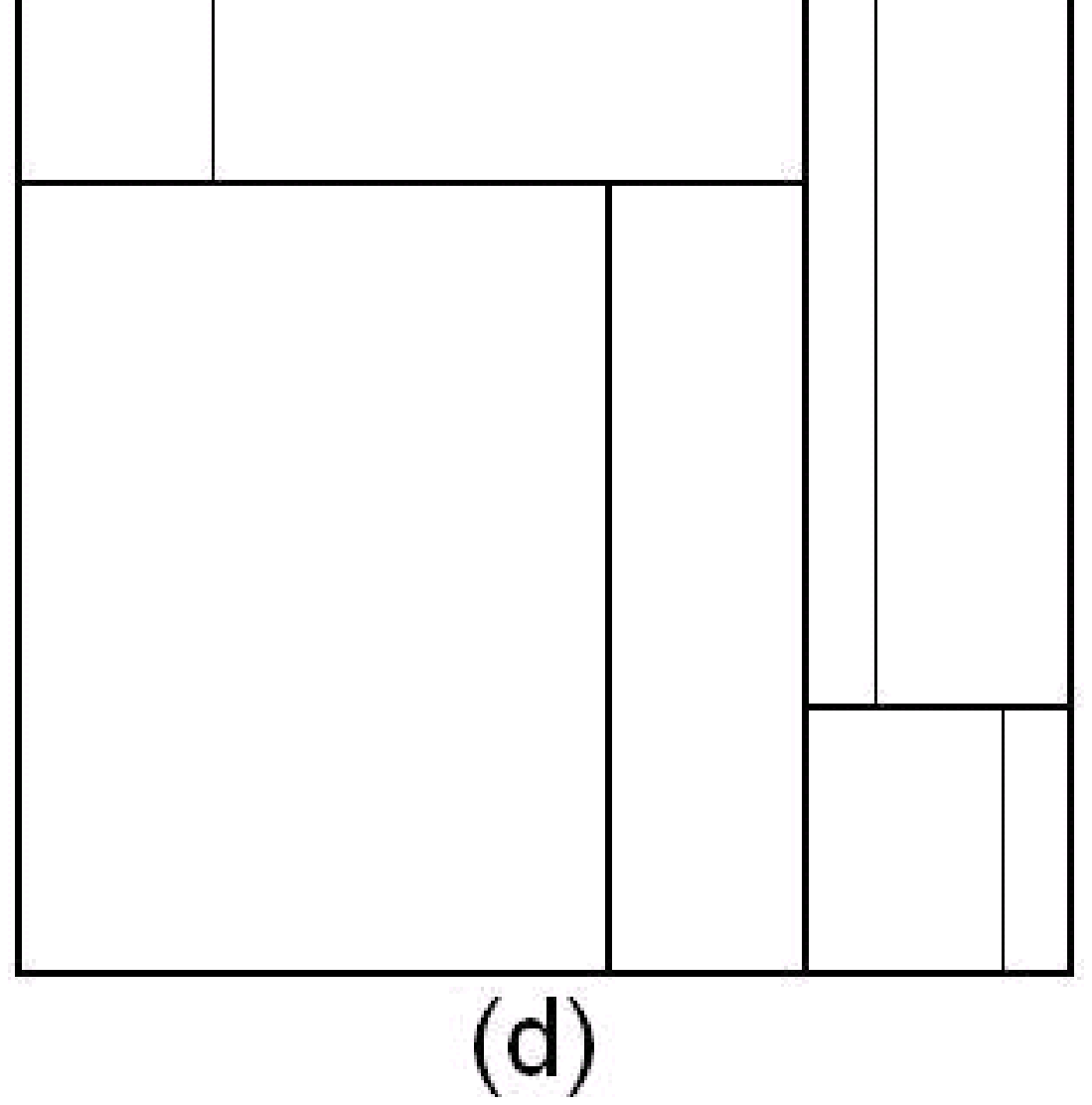}
\includegraphics[width=3cm]{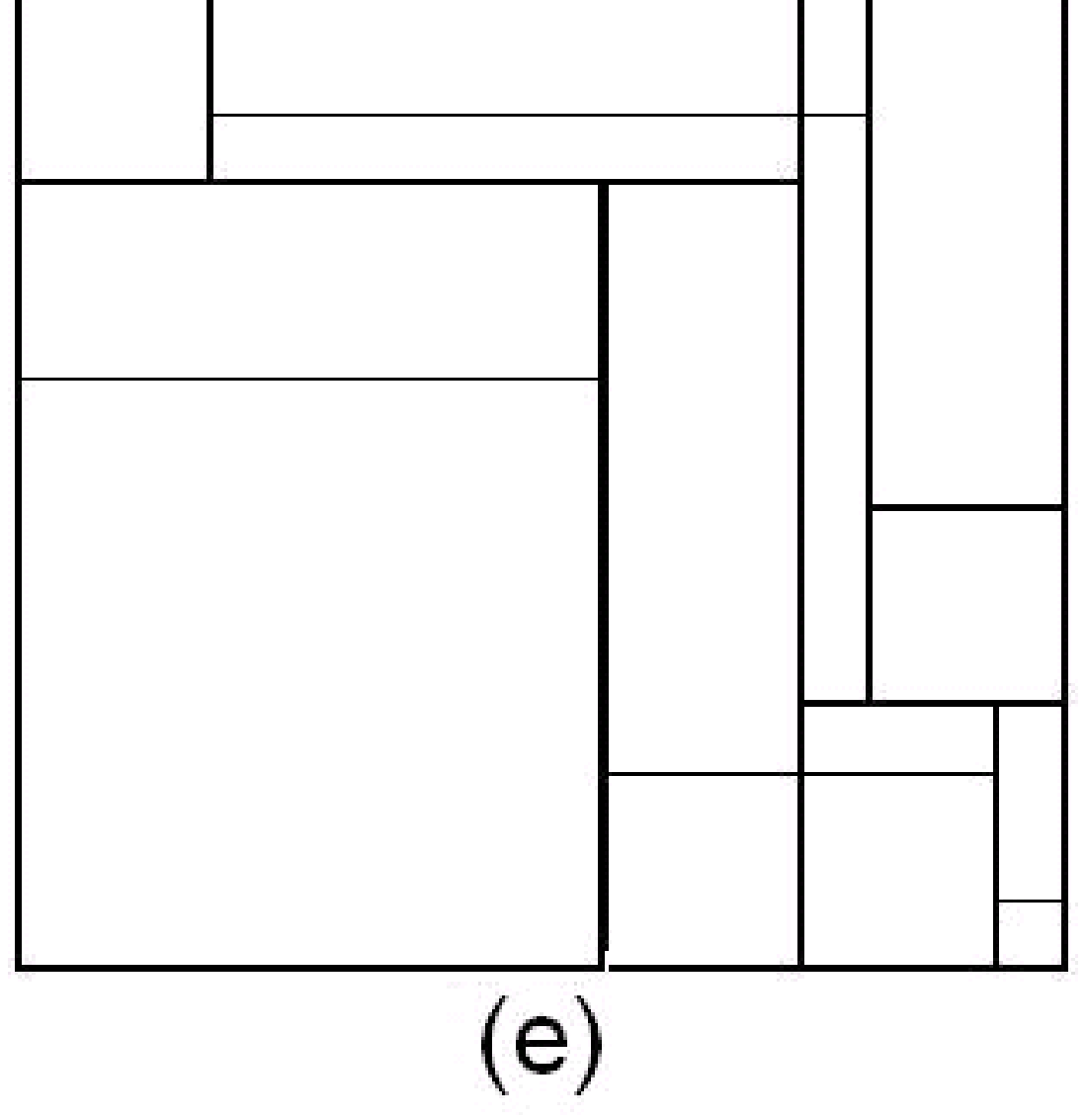}
\caption{ 
\baselineskip=12pt
The first five steps in the construction in the multifractal tiling $Q_{mf}$ 
for $(s,r)=(1,3)$.   }
\label{fig1}
\end{center}
\end{figure}

In Fig.  \ref{fig1} we exemplify the five initial steps of the construction 
of the multifractal for the parameter $\rho=1/3$, or $(s,r)=(1,3)$. 
In \ref{fig1} (a) we show, $n=0$, the 
initial square that by convenience we assume of size $L=1$. 
In (b), $n=1$, a vertical cut is performed and two rectangles are formed. 
We call this a  $(s,r)=(1,3)$ object because  the square is divided in  
$4$ parts such that $1$ part stays at one side and $3$ parts at the 
other side. In (c), $n=2$, two 
horizontal lines are drawn using the same section rate as before. 
At this level the initial square generates
four rectangular blocks. Using as the area unit a square of size 
$\epsilon =1/(s+r)$, the largest block has area $r^2$,
there are two blocks of areas $r s$ and the
smallest block has area $s^2$. 
In (d) and (e), $n=3$ and $n=4$, respectively, 
the same procedure is repeated inside the initial four  blocks. 
In reference \cite{new_phys} it is explored the group of eight possibilities 
of cutting a square lattice with a given ratio.  In this work, as in other 
papers about $Q_{mf}$ \cite{corso1,lim,randi1,randi2,multphy}, 
it is followed the recipe of Fig. \ref{fig1}. 

\begin{figure} \begin{center}
\includegraphics[width=7cm]{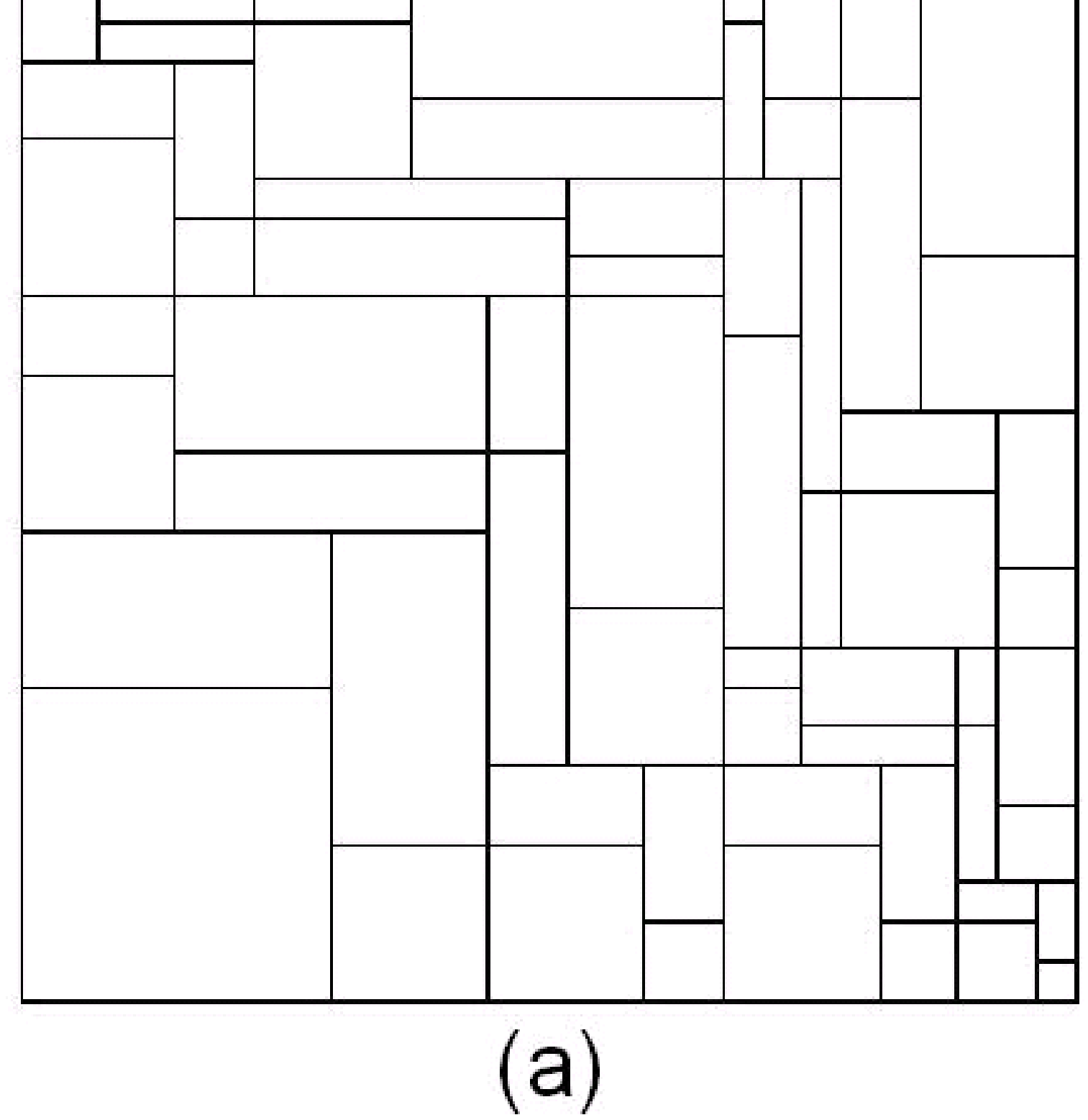}
\includegraphics[width=7cm]{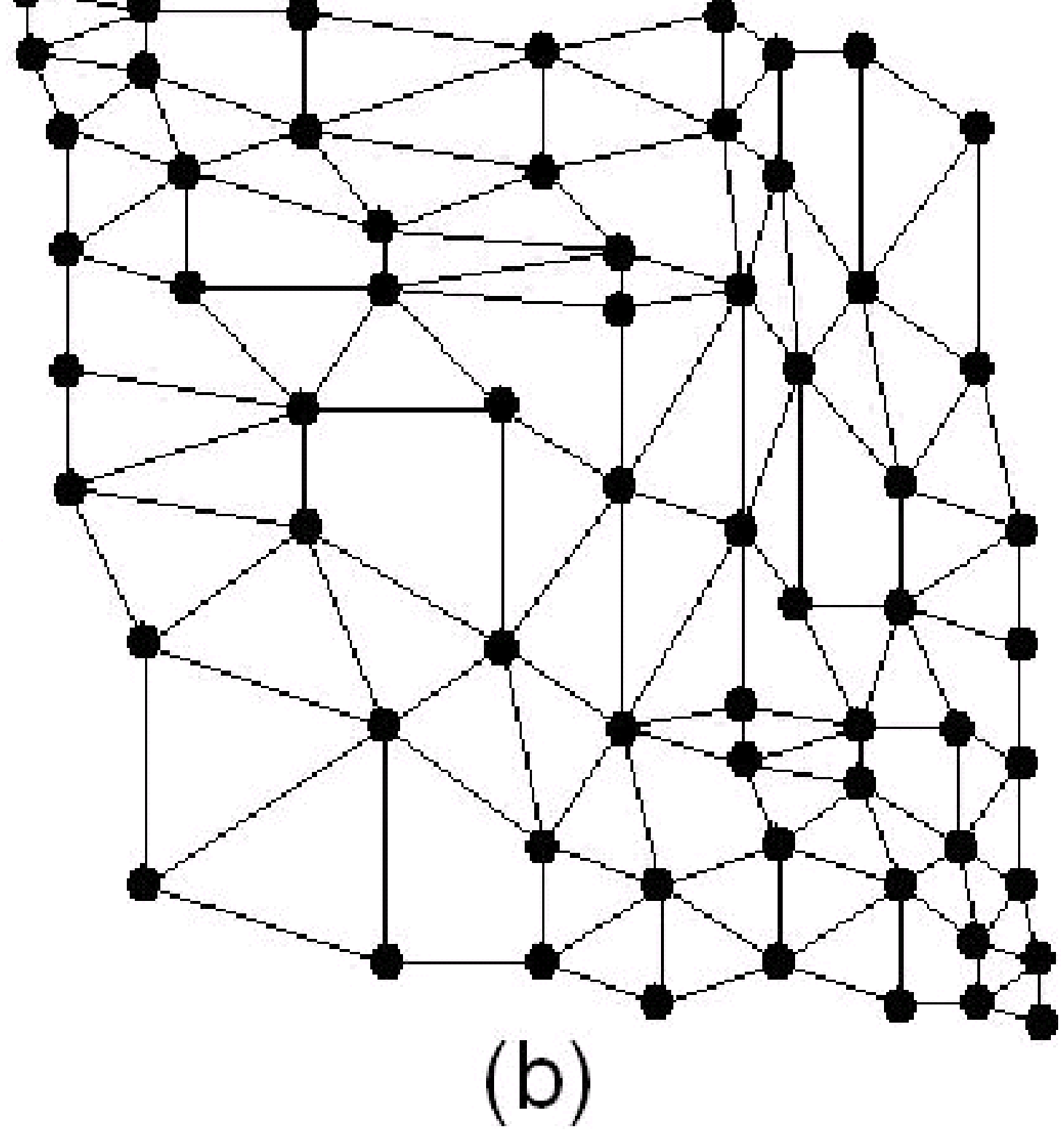} 
\caption{ 
\baselineskip=12pt
A $Q_{mf}$ multifractal tiling and a $Q_{mf}$ network, 
 for $\rho=1/2$ and $n=6$.  
In (a) the original tiling and in (b) the corresponding network. }
\label{fig2}
\end{center}
\end{figure}

We remark that the number of blocks at step $n$ is $2^n$. These blocks do 
not have all the same area, we call the 
subsets of blocks of same area by a $k$-set. 
It is easy to check that the block area distribution 
 follows a binomial distribution and the number of $k$-sets is 
$k=n+1$. This fact implies that the $Q_{mf}$ has the 
remarkable property: in the limit of $n \rightarrow \infty$  
the area of its forming blocks follows a multifractal distribution 
\cite{corso1}.  
The spectrum of fractal distributions comes from a box counting 
reasoning:
\begin{equation}
 D_X= \lim\limits_{\epsilon \rightarrow 0}
          \frac{\log \> N(X)}{\log \> (1/\epsilon)} 
\label{NN}
\end{equation} 
for $N(X)$  the number of unitary cells 
 of size length $\epsilon$ that cover the set of blocks of a 
given area $X$.  Once the initial square is 
partitioned $n$ times,   the size of the unitary cell is 
$\epsilon = 1/(s+r)^{n}$.
For each $k$-set the total area of blocks (using $\epsilon$ area units) is done by:
\begin{equation}
                 N_k = C_n^k \> \> s^k \>r^{(n-k)},
\label{NNN}
\end{equation}
where $C_n^k$ is the binomial coefficient that express
the number of elements $k$-type, and
$s^k \> r^{(n-k)}$ is the area of each element of this set.
 We put together all these elements to have  the fractal dimension of
each $k$-set:
\begin{equation}
                 D_k = \lim\limits_{\epsilon \rightarrow 0} \frac{ \log N_k }{ \log (1/\epsilon)} =
 \lim\limits_{n \rightarrow \infty} \frac{\log (\> C_n^k \>
\> s^k \>r^{(n-k)})}{\log \> (s+r)^n }.
\label{boxcou2}
\end{equation}
This distribution show a concave shape with a maximum at $k=\rho n$. 
The case  $r=s=1$ is degenerated. In this situation the
 subsets of the lattice are composed uniquely by square cells of the same area.
Therefore the tiling is formed by a single subset of dimension $2$. 

In Fig. \ref{fig2} (a) it is shown an example of $Q_{mf}$ construction for 
$(s,r) = (1,2)$ and $n=6$. In Fig.  \ref{fig2} (b) we build a network 
corresponding to this tiling. The nodes of the network are the $2^n$ blocks 
of the $Q_{mf}$ 
and the vertices are established according to a neighborhood criterium. These 
last figures offer a glimpse of the metric heterogeneity and 
the topology of the multifractal. 
In the next section we explore in detail the network properties of this 
class of objects. 

\section{III - Results} 

\begin{figure}
\begin{center}
\includegraphics[width=7cm]{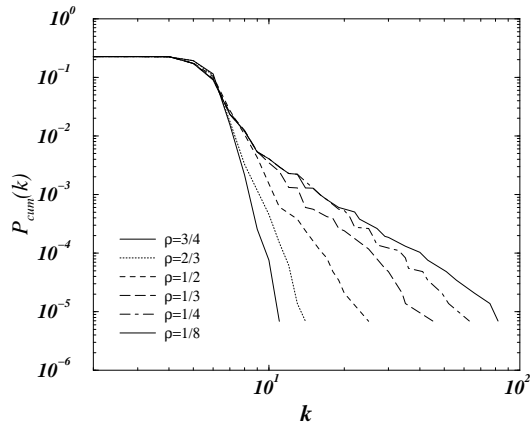}
\caption{
\baselineskip=12pt
The cumulative degree distribution for several values of $\rho$ 
and  $N=2^{16}$. For  $k > 7$ the distributions 
approaches a power-law form. The exponent $\gamma$
controlling the decay of the distributions decreases in the limit of 
$\rho \rightarrow 0$, the full set of exponents is show in Table \ref{tab1}.}
\label{cum}
\end{center}
\end{figure}

We start the analysis of the properties of the $Q_{mf}$ network discussing 
its distribution of connectivity, $P(k)$. In Fig. \ref{cum} we show the cumulative 
sum of $P(k)$ versus $k$ for several values of $\rho$ as indicated in the figure.
 The option for the cumulative sum instead of $P(k)$ itself is 
due to the strong fluctuation of the data. Fig. \ref{cum} confirms the results 
of a previous work \cite{multphy}. For low $k$, typically $k<7$, the 
curve suggests an exponential behavior and above this threshold the network 
depicts a scale-free behavior. The values of the exponents $\gamma$ of 
the power-law $P(k) \propto k^{-\gamma}$ are  indicated in Table \ref{tab1} for 
several $\rho$, trivially the values of $\gamma$ are estimated from 
the slopes of the curve of the cumulative 
sum decreased by one. The exponent $\gamma$ 
goes to an asymptotic limit for large $N$ \cite{multphy}. 
We observe that in the limit of $\rho \rightarrow 1$ the $Q_{mf}$ tiling 
gets more symmetric and at $\rho=1$ the $Q_{mf}$ degenerates into the square 
lattice. For the regular square lattice $P(k)$ is a delta of Dirac centered at $k=4$ 
and the cumulative sum a step function. Fig. \ref{cum} corroborates this 
idea, the  skewed curves in the $\rho \rightarrow 1$ limit anticipate 
the  phase transition at $\rho=1$.

\begin{figure} 
\begin{center}
\includegraphics[width=7cm]{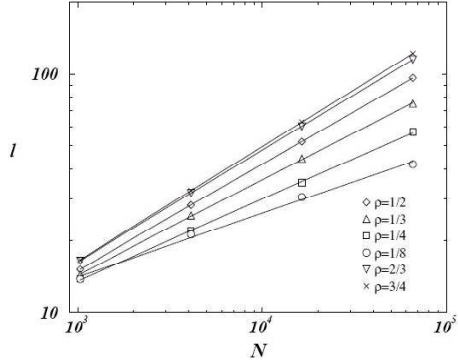}
\caption{ 
\baselineskip=12pt
In (a) it is displayed the average distance
 $\ell$ as a function of N for several values of
$\rho$. }
 \label{fig4}
 \end{center}
 \end{figure}

We explore the distance characterization of the $Q_{mf}$ network in 
 Fig. \ref{fig4} where we display the behavior of the 
average shortest distance for all couple of distinct vertices 
of the network, $\ell$, versus 
network size, $N$. The simulation is performed for some  values of $\rho$ as  
indicated in the figure. For all $\rho$ examined the data show a power-law  
behavior $N \propto \ell^{d_f}$. We find $2 < d_f < 4$, 
the full set of $d_f$ is shown in Table \ref{tab1}.  Two limit cases 
are interesting. The limit $\rho \rightarrow 1$, which corresponds to 
the square lattice, has $d_f \rightarrow 2$ as it is expected in 
a bidimensional space. The opposit limit 
$\rho \rightarrow 0$, which is 
associated with very anisotropic structures, shows large $d_f$. We 
 cannot affirm that $4$ is an asymptotic threshold, further 
numerical investigation should test this hypothesis.   
We remark that the $Q_{mf}$ network does not 
follow a small world relationship $\ell \propto \ln N$ that is common 
to most of  power-law and random like networks.

\begin{figure} 
\begin{center}
\includegraphics[width=7cm]{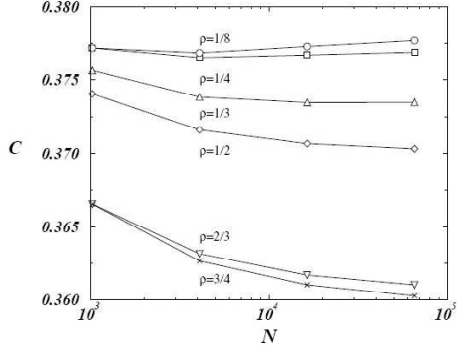} 
\caption{ 
\baselineskip=12pt
In  the clustering coefficient $C$ as function of N for
the same set of values of $\rho$ of the previous figure. }
 \label{fig5}
 \end{center}
 \end{figure}

An analysis of the clustering coefficient, $C$, versus network size, 
$N$, is shown in Fig. \ref{fig5}. The general view of this figure points to a  
stable behavior of $C$ in the limit of large $N$. 
The dispersion of $C$ among $\rho$ is not large, the numerics show 
$C = 0.37 \pm 0.01$. Smaller values of $\rho$, 
however, show a significant larger $C$. 
The discussion about $C$ is intriguing once 
we compare the numerical values of $C$ with the clustering coefficient of a 
random network associated to the $Q_{mf}$ network. An associated random 
network is defined as a network with the same $N$ and $<k>$ of the original 
network (we do not compare our results with a random network with a same 
$P(k)$ because such  random network would alterate the space filling 
characteristics that we are interested in). For a random network  
$C=<k>/N$, in the case of our network: $<k>$ is a constant 
number smaller than $6$ and $N$ a number that can grow without limit. As a 
consequence the associated random network has $C \rightarrow 0$  in the limit  
$N \rightarrow \infty$.  
Therefore the $Q_{mf}$ network has a $C$ that is infinitely larger 
than the clustering coefficient
 of the associated random network. Because of the high $C$ and 
the power-law behavior of distribution of connectivity  we  call the 
$Q_{mf}$ network a complex network.

In Table \ref{tab1} it is shown some parameters related to the 
$Q_{mf}$ network for several values of $\rho$. 
Most of these  data was  already commented in the text.
 We focus now on the average connectivity $< k>$ of the network. 
For all  $\rho$ studied we have $< k> \sim 5.43$ which characterizes  
a sparse network. This is not surprising, since 
there is a result in topology that shows that 
for two dimensions the average coordination number of a tiling
 cannot exceed $6$. 
 The average connectivity confirms Fig. \ref{cum}  
where we can see that the majority of
 vertices are situated in the range: $4\leq k \leq 6$. 
Otherwise we note that most of interesting results concerning the distribution 
of connectivity, in special the power-law behavior, are satisfied only in 
the range $k>7$.  Indeed, the $Q_{mf}$ network, as most of complex networks, also
 have hubs that determine the distinguished characteristics of the network.

\begin{table}[ht]
\caption{The average quantities: clustering coefficient, $C$, minimal distance, $ \ell$, and 
connectivity, $< k> $; the slope  $\gamma$ of the distribution of connectivity 
and the fractal dimension $d_f$. 
The data corresponding to the average parameters
 are  estimated for $N=2^{16}$.}
\vskip 0.5in
\begin{center}
\label{tab1}
\begin{tabular}{|c|c|c|c|c|c|c|}
\hline $\rho$ & $(3,4)$ & $(2,3)$ & $(1,2)$ & $(1,3)$ & $(1,4)$ & $(1,8)$
\\ \hline 
$ C$ & $0.3603$ & $0.3610$ & $0.3703$ & $0.3735$ & $0.3769$ & $0.3777$
\\ \hline
$ \ell$ & $122.02$ & $115.26$ & $96.43$ & $75.12$ & $57.12$ & $41.70$
\\ \hline
$< k > $ & $5.4357$ & $5.4357$ & $5.4338$ & $5.4275$ & $5.4357$ & $5.4357$
\\ \hline
$\gamma$ & $16.6$ & $11.1 $ & $6.4$ & $4.2 $ & $3.4$ & $2.9$
\\ \hline
$d_f$ & $2.09$ & $2.14$ & $2.25$ & $2.52 $ & $2.94$ & $3.79$
\\ \hline
\end{tabular}
\end{center}
\vspace{0.5cm}
\end{table}

\section{IV - Conclusion}

In this work we explore some properties of a space filling network 
that come from a multifractal partition of the square lattice, the 
$Q_{mf}$ network. 
An analysis of the distribution of the connectivity, $P(k)$, assures that the 
$Q_{mf}$ network  shows a power-law tail that is more 
accentuated as increases the anisotropy of the underlying $Q_{mf}$ tiling. 
Roughly the power-law tail of $P(k)$ starts at $k \sim 7$. We remark that 
there is no regular lattice in $2$ dimensions with $k>6$ and typical 
Voronoi latices have an exponential small number of 
vertices in this range. In addition the $Q_{mf}$ network has a clustering 
coefficient that approaches a constant, $C = 0.37 \pm 0.01$,  
that does not depend on $N$. This fact is in contrast to random 
networks that (for a constant $<k>$) have $C \propto N^{-1}$. Because 
the value of $C$ is much larger than the value of $C$ of the  associated 
random network we call the $Q_{mf}$ network a complex network.  

The most interesting aspect of the $Q_{mf}$ network concerns its 
fractal behavior. For the average minimal path $\ell$ we 
observe that $N \propto \ell^{d_f}$ for the fractal dimension, $d_f$.   
The simulations show that $2 < d_f < 4$.  
The lower limit correspond to the 
case $\rho =1$ where the multifractal tiling degenerates 
into the square lattice. In this situation the slope $\gamma$ 
(from the power law $P(k) \propto k^{-\gamma}$) 
increases dramatically, this situation corresponds to the $P(k)$ of 
the square lattice that is of the form of a Delta of Dirac. 
The opposite limit, $\rho =0$, 
corresponding to very anisotropic tilings, presents  comparatively small  
values of $\gamma$.

We point that, diversly from \cite{nature}, 
 the $Q_{mf}$ network shows an actual fractal behavior 
$N \propto \ell^{d_f}$ that is obtained without any renormalization 
artefact. In the reference \cite{nature} a ingenious procedure 
is used to calculate two fractal dimensions $d_B$ and $d_f$. The 
first dimension depends on a suitable embedding in a metric space and a box counting 
methodology. The second is based on network distance and a mass (number 
of vertices) inside a given radios. In our case, we have calculated $d_f$ in 
the standard way, the box-counting, however, depends on the methodology 
we use to make the embedding of the $Q_{mf}$ object. The simplest embedding 
is the $Q_{mf}$ lattice itself that is a 2-dimensional object and as a result 
$d_B=2$. Note that the multifractal property of $Q_{mf}$ appears when we 
consider the subsets of blocks of same area, if we disregard the 
area set a bidimensional tilling assumes the trivial topologic dimension, $d_f=2$.

The $Q_{mf}$ tiling is indeed a remarkable mathematical object, from a 
metric perspective it is a multifractal: it is formed by a 
denumerable quantity of sets of different areas each one with 
a given fractal dimension. In a topologic perspective the
connections among the vertices (the cells of the tiling) 
form a fractal network. We remark that regular networks 
(generated from lattices for instance) and the  Bethe tree satisfy the 
criterium $N \propto \ell^{d}$,  
but these  structures are regular.  
For the best knowledgement of the 
authors the $Q_{mf}$ network is the only case of a 
true fractal, scale-free and with high clustering coefficient.

\vspace{1cm}

\centerline{\bf Acknowledgments}

The authors gratefully acknowledge the financial support of Conselho Nacional
de Desenvolvimento Cient{\'\i}fico e Tecnol{\'o}gico (CNPq)-Brazil,
FINEP and Programa PET-SESU/MEC. D. J. B. S. Thanks to D. R. de Paula.

\end{document}